\documentclass[aps,twocolumn,groupedaddress,showpacs]{revtex4}

\usepackage{graphicx}

\begin{document}

% You should use BibTeX and revtex.bst for references
\bibliographystyle{apsrev}

\title{
Effect of a magnetic field on the long-range magnetic order 
in insulating Nd$_{2}$CuO$_{4}$, nonsuperconducting and superconducting 
Nd$_{1.85}$Ce$_{0.15}$CuO$_{4}$\\}

\author{M. Matsuura,$^1$ Pengcheng Dai,$^{2,1,\ast}$ 
H. J. Kang,$^2$ J. W. Lynn,$^3$ D. N. Argyriou,$^4$ 
K. Prokes,$^{4,5}$ Y. Onose,$^6$ and Y. Tokura$^{6,7,8}$ 
}

\address{$^1$Condensed Matter Sciences Division, 
Oak Ridge National Laboratory, Oak Ridge, Tennessee 37831-6393 }

\address{$^2$Department of Physics and Astronomy, 
The University of Tennessee, Knoxville, Tennessee 37996-1200}

\address{$^3$NIST Center for Neutron Research, 
National Institute of Standards and Technology, 
Gaithersburg, Maryland 20899}

\address{$^4$Hahn-Meitner Institute, Glienicker Str 100, 
Berlin D-14109, Germany}

\address{$^5$Department of Electronic Structures, 
Charles University, Ke Karlovu 5, 12116 Prague 2, 
The Czech Republic}

\address{$^6$Spin Superstructure Project, ERATO, 
Japan Science and Technology, Tsukuba 305-8562, Japan}

\address{$^7$Correlated Electron Research Center, 
Tsukuba 305-8562, Japan}

\address{$^8$Department of Applied Physics, 
University of Tokyo, Tokyo 113-8656, Japan}

\date{\today}

\begin{abstract}
We have measured the effect of a $c$-axis aligned magnetic field 
on the long-range magnetic order of insulating Nd$_{2}$CuO$_{4}$, as-grown
nonsuperconducting 
and superconducting Nd$_{1.85}$Ce$_{0.15}$CuO$_{4}$. 
On cooling from room temperature, 
Nd$_{2}$CuO$_{4}$ goes through a series of antiferromagnetic (AF) 
phase transitions with different noncollinear spin structures. 
In all phases of Nd$_{2}$CuO$_{4}$, 
we find that the applied $c$-axis field induces 
a canting of the AF order but does not alter 
the basic zero-field noncollinear spin structures. Similar 
behavior is also found in as-grown nonsuperconducting 
Nd$_{1.85}$Ce$_{0.15}$CuO$_{4}$. 
These results contrast dramatically 
with those of superconducting Nd$_{1.85}$Ce$_{0.15}$CuO$_{4}$, 
where a $c$-axis aligned magnetic field induces a static, 
anomalously conducting, long-range ordered AF state. 
We confirm that the annealing process necessary to make 
superconducting Nd$_{1.85}$Ce$_{0.15}$CuO$_{4}$ also induces 
epitaxial, three-dimensional long-range ordered cubic (Nd,Ce)$_2$O$_3$ as an 
impurity phase. In addition, the annealing process makes a series of  
quasi two-dimensional superlattice reflections associated with lattice distortions of  
Nd$_{1.85}$Ce$_{0.15}$CuO$_{4}$ in the CuO$_2$ plane.
While the application of a magnetic field will induce a net moment in the
impurity phase, we determine its magnitude 
and eliminate this as a possibility for the observed 
magnetic field-induced effect in superconducting Nd$_{1.85}$Ce$_{0.15}$CuO$_{4}$.
\end{abstract}

\pacs{74.72.Jt, 75.25.+z, 75.50.Ee, 61.12.Ld}

\maketitle
\narrowtext
\section{Introduction}
High-transition-temperature (high-$T_c$) superconductivity 
occurs in lamellar copper oxides when 
holes \cite{bednorz} 
or electrons \cite{tokura,takagi} are doped into the CuO$_2$ planes. 
For the parent compounds of hole-doped materials 
such as La$_2$CuO$_4$ and YBa$_2$Cu$_3$O$_6$,
the Cu$^{2+}$ spins order at relatively high temperatures 
($\sim$300 K and 420 K, respectively) in a simple
antiferromagnetic (AF) collinear structure that 
doubles the crystallographic unit cell in the CuO$_2$ planes \cite{vaknin,tranquada}. 
Although the parent compound of electron-doped copper oxides such as Nd$_{2}$CuO$_{4}$
also has AF spin structures doubling the CuO$_2$ unit cell, the Cu$^{2+}$ moments 
order in three phases with two different AF noncollinear spin structures as shown in 
Figs. 1(a) and 1(b) \cite{skanthakumar0,jeff,skanthakumar1,skanthakumar2}. 
These noncollinear spin structures appear in Nd$_{2}$CuO$_{4}$
 because of the presence of the magnetic exchange interaction between 
Cu$^{2+}$ and Nd$^{3+}$ \cite{sachidanandam,petitgrand}.
Compared to the hole-doped 
La$_{2-x}$Sr$_{x}$CuO$_4$,
the long-range AF order in electron-doped Nd$_{2-x}$Ce$_{x}$CuO$_4$  
persists to much larger $x$ ($\geq 0.12$) \cite{tokura,takagi}, and coexists 
with superconductivity for even the highest $T_c$ (= 25 K) material ($x=0.15$) \cite{yamada,uefuji}. 
In contrast, superconductivity in La$_{2-x}$Sr$_{x}$CuO$_4$ 
emerges from a spin-glass regime and occurs over a wider 
doping concentration.

The close proximity of AF order and superconductivity raises an interesting question
concerning the role of long-range magnetic order in the superconductivity of copper oxides.    
Theoretically, it was predicted that, when an applied field creates vortices in these
superconductors, AF order would be induced in the core of each vortex 
\cite{zhang,arovas}. For underdoped superconducting La$_{2-x}$Sr$_x$CuO$_4$, 
neutron scattering experiments  
show that  a $c$-axis aligned 
magnetic field (${\bf B}||c$-axis) not only suppresses superconductivity but also enhances 
the static incommensurate spin density wave order,
thus suggesting that such order competes directly with 
superconductivity \cite{katano,lake1,lake2,khaykovich}.
Although muon spin resonance ($\mu$SR) \cite{miller} and nuclear
magnetic resonance (NMR) experiments in underdoped YBa$_2$Cu$_3$O$_{6.5}$
\cite{mitrovic} also suggest enhanced AF order originating from regions
near the vortex core, neutron scattering experiments failed to confirm 
any enhancement of the static long-range order in 
YBa$_2$Cu$_3$O$_{6.6}$ for fields up to 7-T \cite{dai,mook}. 
Therefore, in spite of 
intensive effort \cite{katano,lake1,lake2,khaykovich,miller,mitrovic,dai,mook}, the nature of the 
superconductivity-suppressed ground state in high-$T_c$ superconductors 
is still unknown.

The major difficulty in studying the ground state of hole-doped
high-$T_c$ superconductors is the enormous 
upper critical fields $B_{c2}$ ($>20$-T)   
required to completely suppress superconductivity. Fortunately, electron-doped materials
generally have $B_{c2}$, for magnetic fields
aligned along the $c$-axis, less than 10-T \cite{hidaka,fournier,hill2001,wang2003}, a 
value easily reachable in neutron scattering experiments. 
While recent experiments by Matsuda {\it et al.} \cite{matsuda} found
that a 10-T $c$-axis aligned field has no effect on the AF order in 
the superconducting Nd$_{1.86}$Ce$_{0.14}$CuO$_4$, we showed that 
such fields in 
Nd$_{1.85}$Ce$_{0.15}$CuO$_4$ enhance the AF moment and 
induce a direct quantum phase transition from the superconducting state 
to an anomalously conducting antiferromagnetically ordered state at $B_{c2}$ \cite{kang}.
The induced AF moments scale approximately linearly with the applied field,
 saturate at $B_{c2}$, and then decrease for higher fields, 
indicating that the field-induced AF order competes directly  
with superconductivity \cite{kang}.

Although electron-doped Nd$_{2-x}$Ce$_{x}$CuO$_4$
offers a unique opportunity for studying the superconductivity-suppressed ground state of 
high-$T_c$ copper oxides, the
system is somewhat more complicated than hole-doped materials such as La$_{2-x}$Sr$_x$CuO$_4$
and YBa$_2$Cu$_3$O$_{6+x}$
for three reasons. 
First, it contains two magnetic ions (rare-earth Nd$^{3+}$ and Cu$^{2+}$), and the ordered
Cu sublattice induces the long-range AF ordering of Nd ions \cite{lynnprb}. 
The effect of an applied field on rare-earth Nd$^{3+}$ magnetic moments and their ordering 
is unknown. Second, 
for even the highest $T_c$ (= 25 K) Nd$_{2-x}$Ce$_{x}$CuO$_4$ ($x=0.15$), 
superconductivity coexists with 
the long-range residual AF order, and the nature of their coexistence is unclear \cite{yamada,uefuji}. 
Finally, superconductivity in Nd$_{2-x}$Ce$_{x}$CuO$_4$ can only be achieved by annealing
the as-grown samples at high temperatures \cite{yamada,uefuji,onose}. The annealing 
process not only induces superconductivity in 
Nd$_{2-x}$Ce$_{x}$CuO$_4$, but also produces structural superlattice reflections 
of unknown origin \cite{kurahashi}.

To understand the effect of a magnetic field on superconductivity in Nd$_{1.85}$Ce$_{0.15}$CuO$_4$ \cite{kang},
one must first determine its influence on the residual AF order 
without the complication of superconductivity. 
Since the residual AF order 
in superconducting Nd$_{2-x}$Ce$_{x}$CuO$_4$ 
has the same magnetic structure as that of the insulating
Nd$_2$CuO$_4$ at low temperatures  \cite{uefuji}, 
investigating the field effect on AF orders in Nd$_{2}$CuO$_{4}$ 
will resolve this issue. Second, the effect of Ce doping in Nd$_{2}$CuO$_{4}$
can be studied by performing magnetic field experiments in as-grown nonsuperconducting 
Nd$_{1.85}$Ce$_{0.15}$CuO$_4$. Finally, to resolve the nature of the coexisting 
superconducting and AF orders in
Nd$_{2-x}$Ce$_{x}$CuO$_4$, we also need to understand the microscopic origin of the
superlattice reflections and the
effect of a magnetic field on these reflections.

This article describes experiments designed to understand the effect of a $c$-axis aligned 
magnetic field in all noncollinear spin structure phases of Nd$_{2}$CuO$_{4}$ and in 
residual AF order of as-grown nonsuperconducting and 
superconducting Nd$_{1.85}$Ce$_{0.15}$CuO$_4$. For Nd$_{2}$CuO$_{4}$, 
previous work showed that a magnetic
field applied parallel to the CuO$_2$ planes transforms the spins from the noncollinear 
to collinear AF structure \cite{skanthakumar1,skanthakumar2}. We find that a
field applied perpendicular to the CuO$_2$ planes only induces a canting
of the AF moment, and does not change the noncollinear nature of spin structures in
all phases of Nd$_{2}$CuO$_{4}$. 
For nonsuperconducting Nd$_{1.85}$Ce$_{0.15}$CuO$_4$,
a 7-T $c$-axis aligned field does not enhance the AF moment at low temperature. 
Finally, we confirm that the annealing process necessary to make 
superconducting Nd$_{1.85}$Ce$_{0.15}$CuO$_{4}$ also induces 
epitaxial, three-dimensional ordered cubic (Nd,Ce)$_2$O$_3$ (space group $Ia3$, and lattice
parameter $a_{NO}=11.072$ \AA) as an impurity phase \cite{meng,brauer}. 
In addition, the annealing process makes a series of  
quasi two-dimensional superlattice reflections associated with lattice distortions of  
Nd$_{1.85}$Ce$_{0.15}$CuO$_{4}$. While these quasi two-dimensional superlattice reflections
have no field-induced effect, we determine the field-induced effect in the impurity phase  
and show that such effect cannot account for the anisotropy in field-induced intensity  
between ${\bf B}||c$-axis and ${\bf B}||ab$-plane.
To further demonstrate that field-induced scattering is an intrinsic 
property of the superconductor,
we probed Bragg reflections exclusively
from Nd$_{1.85}$Ce$_{0.15}$CuO$_{4}$ by performing experiments 
using a horizontal field magnet. The results confirm 
that AF signal arises from the suppression of 
superconductivity by the $c$-axis aligned field 
in Nd$_{1.85}$Ce$_{0.15}$CuO$_4$ \cite{kang}.

\section{Experimental Details}

Our experiments were performed on the BT-2 and BT-9 triple-axis
spectrometers at the NIST Center for Neutron Research and
on the E4 two-axis diffractometer at the Berlin Neutron Scattering Center, Hahn-Meitner-Institute (HMI). 
We measure momentum transfer $(q_x,q_y,q_z)$ in units of \AA$^{-1}$ and specify 
the reciprocal space positions in reciprocal lattice units (rlu) 
$(H,K,L)=(q_xa/2\pi,q_ya/2\pi,q_zc/2\pi)$ appropriate for the tetragonal unit cells of
Nd$_{2}$CuO$_{4}$ (space group $I4/mmm$, $a=3.944$ and $c=12.169$ \AA) and Nd$_{1.85}$Ce$_{0.15}$CuO$_4$ 
(space group $I4/mmm$, $a=3.945$ and $c=12.044$ \AA), where $a$ and $c$ are in-plane and out-of-plane
lattice parameters, respectively.

\begin{figure}
\includegraphics[keepaspectratio=true,width=0.8\columnwidth,clip]{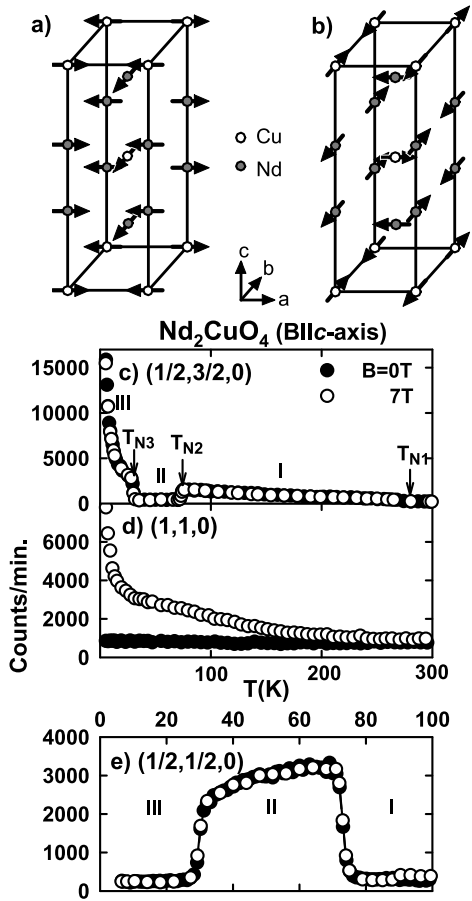}
\caption{Spin structure models and temperature dependent 
scattering at AF and ferromagnetic (FM) Bragg positions in Nd$_{2}$CuO$_{4}$.
The schematic diagrams show the non-collinear spin structures in 
(a) type-I ($75<T<275$ K) and type-III ($T<30$ K) phases,
and (b) type-II ($30<T<75$ K) phase.
The closed and open circles in (c)-(e) 
represent the scattering intensity at $B=0$ and 7-T,
respectively, for fields aligned along the $c$-axis (${\bf B}||c$-axis).
}
\end{figure}

For NIST experiments, the collimations were, proceeding from the reactor to the detector,
40$^\prime$-46$^\prime$-sample-40$^\prime$-80$^\prime$ (full-width at half-maximum), and the final
neutron energy was fixed at $E_{f}=14.7$ meV. The monochromator, analyzer and filters were all
pyrolytic graphite. We aligned the CuO$_2$ planes in the horizontal $[H,K,0]$ scattering plane and 
applied the vertical magnetic field along the $c$-axis (${\bf B}||c$-axis). 
In this geometry, we can access reciprocal space at any $(H,K,0)$. 
To determine anisotropy of the field-induced effect, we also performed experiments in the
$(H,H,L)$ scattering plane where the applied vertical fields are along 
the $[1,-1,0]$ direction (${\bf B}||ab$-plane).   
For E4 measurements at HMI, we used 40$^\prime$-40$^\prime$-sample-40$^\prime$ 
collimation with fixed incident neutron energy of $E_i=13.6$ meV. 
A pyrolytic graphite filter was placed in front of the sample to
eliminate higher-order contamination.
The HM-2 4-T horizontal field magnet was used 
to apply a $c$-axis aligned field while probing the $L$ modulation of the scattering.
Although these measurements are crucial in determining the field-induced magnetic structure,
the highly restricted access angles of the magnet limit the regions of reciprocal space
that can be probed. In horizontal field measurements on Nd$_{1.85}$Ce$_{0.15}$CuO$_4$, the crystal
was aligned in the $(H,H,L)$ zone  and the applied field was along the $c$-axis.

We grew a single crystal of Nd$_{2}$CuO$_{4}$ ($\phi 7\times 20$ mm) and crystals of 
Nd$_{1.85}$Ce$_{0.15}$CuO$_4$ 
using the traveling solvent
floating zone technique \cite{onose}. 
The Nd$_{2}$CuO$_{4}$ crystal used in the experiments is as-grown.
We also performed experiments on as-grown nonsuperconducting Nd$_{1.85}$Ce$_{0.15}$CuO$_4$
and superconducting Nd$_{1.85}$Ce$_{0.15}$CuO$_4$.
Superconductivity in Nd$_{1.85}$Ce$_{0.15}$CuO$_4$ was obtained after annealing the samples in a flowing Ar/O$_2$
gas mixture with a partial oxygen pressure of $\sim 10^{-5}$ ATM at 1000$^\circ$C for 100 h.
Magnetic susceptibility 
measured on small pieces of crystals ($\sim200$ mg) cut from the samples used for neutron experiments
show the onset of bulk superconductivity at $T_c\approx$25 K with a transition width of 3 K.
With a $c$-axis aligned field of 4 Oe, the zero-field cooled data show complete screening of flux.
In the field cooled case, the crystal expels 18\% of the flux, indicating 
that the bulk
superconductivity has at least 18\% of the volume fraction. 
The susceptibility of Nd$_{2-x}$Ce$_x$CuO$_{4}$ 
in the CuO$_2$ planes is several times bigger than that perpendicular to them.
The large 
magnetic anisotropy means that a $c$-axis aligned field acting on the magnetic moments (Nd and Cu)
produces a large torque on the sample. To prevent the samples from rotating under the influence
of a ${\bf B}||c$-axis field, they were clamped on solid aluminum
brackets. For experiments at NIST,
the bracket was inserted inside a He filled aluminum can mounted on a standard
7-T split-coil superconducting magnet. For HMI experiments, the sample assembly 
was mounted on a mini-goniometer and inserted directly to the sample chamber of the HM2
magnet. 

In an attempt to determine the homogeneity of the annealed superconducting
and as-grown nonsuperconducting 
Nd$_{1.85}$Ce$_{0.15}$CuO$_4$, 
we performed neutron diffraction measurements on both samples. The results 
confirm that superlattice 
reflections exist only in superconducting samples \cite{kurahashi},  
and are resolution limited   
(indicating a correlation length larger than 300 \AA) in the CuO$_2$ plane 
and relatively broad along the $c$-axis (see sections IV-V). 
Although the sharpness of the superlattice reflections 
and susceptibility measurements suggest that the crystal is homogeneous with
bulk superconductivity, it is not clear how the residual AF order, superstructure,
and superconductivity coexist microscopically in the material.

\section{Results on insulating N\lowercase{d}$_{2}$C\lowercase{u}O$_{4}$}

Before describing the field effect on the long-range magnetic order
of Nd$_{2}$CuO$_{4}$, we briefly review its zero-field behavior.
As shown in Refs. \cite{skanthakumar0,jeff,skanthakumar1,skanthakumar2},
the Cu spins in 
Nd$_{2}$CuO$_{4}$ first order into the noncollinear type-I spin 
structure below $T_{N1}=275$ K (Fig. 1a). On further cooling, 
the Cu spins re-orient into type-II (at $T_{N2}=75$ K) and type-III ($T_{N3}=30$ K) phases.
In the type-II phase (Fig. 1b), all the Cu spins rotate by 90$^\circ$ about the $c$-axis 
from the type-I phase. They rotate back to their original direction below $T_{N3}$ 
in the type-III phase (Fig. 1a). The closed circles in Figs. 1c, 1d, and 1e 
show the temperature dependence of the scattering at $(1/2,3/2,0)$, $(1,1,0)$,
and $(1/2,1/2,0)$, respectively. Clear AF phase transitions are seen at 
$T_{N1}$, $T_{N2}$, and $T_{N3}$ as marked by the arrows in Fig. 1c,
confirming previous work  \cite{skanthakumar0,jeff,skanthakumar1,skanthakumar2}. 
The large intensity increase of the $(1/2,3/2,0)$ peak
below $\sim$20 K is associated with staggered moments
on Nd sites induced by Cu-Nd coupling. Magnetic 
structure factor calculations indicate that the $(1/2,1/2,0)$ reflection
has vanishing intensity in the type-I/III phases and becomes finite 
in the type-II phase. The large intensity jumps of $(1/2,1/2,0)$  at
$T_{N2}$ and $T_{N3}$ shown in Fig. 1e clearly bear this out. 
Since Nd$_{2}$CuO$_{4}$ only has AF phase transitions at zero-field, 
the intensity of the nuclear Bragg peak $(1,1,0)$ has no magnetic contributions and hence
is essentially temperature independent (Fig. 1d). 
To estimate the magnetic moments of Nd and
Cu in different phases of Nd$_{2}$CuO$_{4}$, we normalized the intensity of AF peaks
at $(1/2,1/2,0)$ and $(1/2,3/2,0)$ to that of the weak $(1,1,0)$ or strong $(2,0,0)$ nuclear Bragg peak.
The estimated Nd and Cu moments differ dramatically depending on the chosen Bragg peaks,
dure primarily to extinction but also Nd absorption of the large crystal (Table I). In particular,
the intensities of the strong peaks such as $(2,0,0)$ are 
severely extinction limited, which overestimates the 
magnitude of the ordered moment.

\begin{table}
\caption{The magnitude of magnetic moments calculated by normalizing the AF 
intensity at $(1/2,1/2,0)$ and $(1/2,3/2,0)$ to that of the weak $(1,1,0)$ 
or strong $(2,0,0)$ nuclear Bragg reflection. From the powder diffraction 
measurements on Nd$_2$CuO$_4$ \cite{jeff}, the ordered Cu moment was estimated to be 0.46 $\mu_B$ 
at 80 K and Nd moment was 0.46 $\mu_B$ at 5 K. In computing the Cu and Nd moments,
we assumed that the Nd moment does not contribute to magnetic scattering above 50 K and the  
Cu moment does not change below 50 K.}
\begin{ruledtabular}
\begin{tabular}{ccccc}
$(H,K,L)$ & $T$(K) & 5 & 55 & 100 \\
\hline
$(1,1,0)$ & $M_{Cu}$ ($\mu_B$) & 0.1           & $0.1\pm0.01$ & $0.12\pm0.04$ \\
& $M_{Nd}$ ($\mu_B$) & $0.16\pm0.01$ & 0           & 0 \\
$(2,0,0)$ & $M_{Cu}$ ($\mu_B$) & 1.2           & $1.2\pm0.2$ & $1.0\pm0.2$ \\
\end{tabular}
\end{ruledtabular}
\end{table}

\begin{figure}
\includegraphics[keepaspectratio=true,width=0.8\columnwidth,clip]{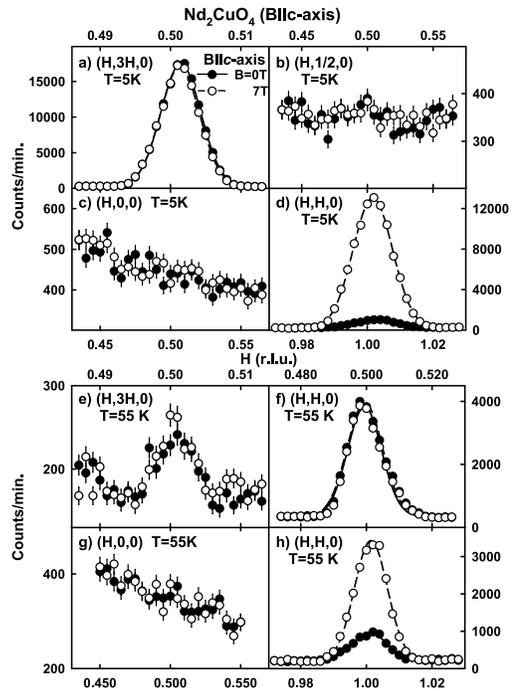}
\caption{Effect of a ${\bf B}||c$-axis field on the AF peaks (half integer)
and field-induced FM peaks in the type-III (a-d)
and type-II (e-h) AF phases of Nd$_{2}$CuO$_{4}$. 
Scans around (a),(e) AF Bragg reflections 
$(1/2,3/2,0)$, (b),(f) $(1/2,1/2,0)$, 
(c),(g) $(1/2,0,0)$ (position of the field-induced scattering
observed in superconducting Nd$_{1.85}$Ce$_{0.15}$CuO$_{4}$), 
and (d),(h) the FM Bragg peak $(1,1,0)$
at 5 K (type-III phase) and 55 K (type-II phase).
The closed and open circles 
represent identical scans at zero and 7-T field,
respectively. The $q$-width of the zero field and field-induced FM scattering
are resolution-limited and identical, thus implying an in-plane correlation
length larger than 300 \AA.
The solid and dotted lines are Gaussian fits.}
\end{figure}

For superconducting Nd$_{1.85}$Ce$_{0.15}$CuO$_4$, 
a ${\bf B}||c$-axis field induces long-range ordered AF peaks at low temperatures that
obey the selection rules $(\pm(2m+1)/2,\pm(2n+1)/2,0)$, $(\pm(2m+1)/2,\pm n,0)$, and
$(\pm m,\pm(2n+1)/2,0)$ with $m, n = 0, 1, 2$ \cite{kang}. 
While magnetic peaks at $(1/2,0,0)$ and $(1/2,1/2,0)$  
are purely field-induced and not present in zero field, 
$(1/2,3/2,0)$ type reflections associated with 
the zero-field AF order (Fig. 1a) are also enhanced \cite{yamada,uefuji}.
Furthermore, AF order appears to saturate at $B_{c2}$ 
while the FM intensity at $(1,1,0)$
continues to rise for fields above $B_{c2}$ \cite{kang}.
To see if a ${\bf B}||c$-axis field can also induce 
magnetic peaks around $(1/2,0,0)$ and $(1/2,1/2,0)$ without the presence of
superconductivity, we performed 
experiments at $T=5$ K ($<T_{N3}$) where Nd$_{2}$CuO$_{4}$
has the identical (type-III phase) spin structure as that of 
Nd$_{1.85}$Ce$_{0.15}$CuO$_4$  at $T=5$ K (see Figs. 1a and 1c).
No signal was observed.

Figures 2 shows 
scans around the AF positions $(1/2,3/2,0)$,
$(1/2,1/2,0)$, $(1/2,0,0)$, and structural Bragg reflection 
$(1,1,0)$. At zero field and 5 K (closed circles in Figs. 2a), 
we observe a resolution-limited magnetic peak at $(1/2,3/2,0)$ 
as expected from the type-III spin structure. 
However, scans around $(1/2,1/2,0)$ and $(1/2,0,0)$ (Figs. 2b and 2c) 
show no evidence of the weak structural superlattice
peaks seen in the superconducting Nd$_{1.85}$Ce$_{0.15}$CuO$_4$ sample \cite{kang,kurahashi}.
The sloping background along the $[H,0,0]$ direction around 
$(1/2,0,0)$ is due to the small scattering angles for this scan.
On application of a 7-T ${\bf B}||c$-axis field, long-range FM ordering
is induced as seen by the added magnetic intensity to the $(1,1,0)$ structural
Bragg peak intensity (Fig. 2d). Such enhancement is most likely due to the polarization
of the Nd moment in the sample. 
On the other hand, the lack of intensity changes 
between 0 and 7-T data at $(1/2,3/2,0)$, $(1/2,1/2,0)$, and 
$(1/2,0,0)$ positions (Figs. 2a-c)  
indicate that the applied field neither enhances the type-III AF order nor 
induces a new AF state. In contrast, a 7-T ${\bf B}||c$-axis field induces
magnetic scattering at all these positions below $T_c$ 
in superconducting Nd$_{1.85}$Ce$_{0.15}$CuO$_4$ \cite{kang}.

To determine the effect of a 7-T field on the type-II phase, we repeated
the measurements around $(1/2,3/2,0)$, $(1/2,1/2,0)$, 
$(1/2,0,0)$, and $(1,1,0)$ positions at 55 K. The outcome of the experiments 
plotted in Figs. 2e-h clearly shows that a 7-T field only 
induces FM ordering at $(1,1,0)$ and has negligible effect on the intensities of 
type-II AF Bragg reflections. Comparing Figs. 2e-h with Figs. 2a-d, we find that 
the FM enhancement of the $(1,1,0)$ reflection 
is smaller in the type-II phase at $T=55$ K, and the 
$(1/2,1/2,0)$ reflection that is forbidden in the type-III spin
structure becomes visible.

\begin{figure}
\includegraphics[keepaspectratio=true,width=0.8\columnwidth,clip]{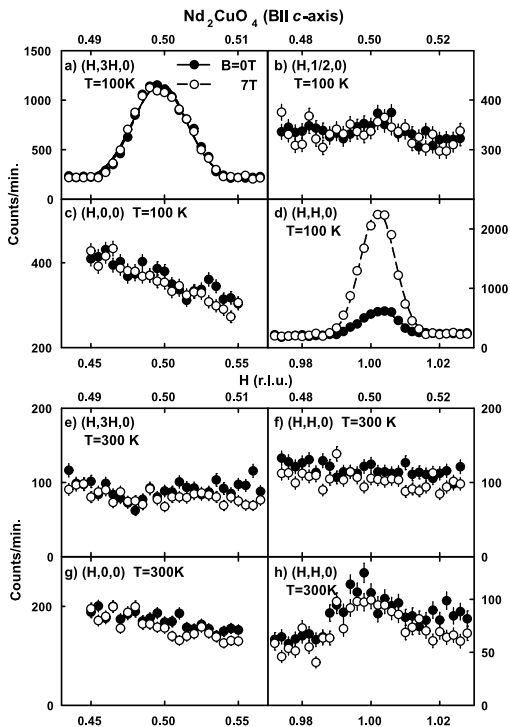}
\caption{Effect of a ${\bf B}||c$-axis field on the AF peaks (half integer)
and field-induced FM peaks in the AF type-I phase (a-d)
and paramagnetic (e-h) state of Nd$_{2}$CuO$_{4}$. 
Scans around (a),(e) AF Bragg reflections 
$(1/2,3/2,0)$, (b),(f) $(1/2,1/2,0)$, 
(c),(g) $(1/2,0,0)$ (position of the field-induced scattering
observed in superconducting Nd$_{1.85}$Ce$_{0.15}$CuO$_{4}$), 
and (d),(h) the FM Bragg peak $(1,1,0)$
at 100 K (type-I phase) and 300 K (paramagnetic state).
The closed and open circles
represent identical scans at zero and 7-T field,
respectively. Since no peaks are observed at half
integer positions in the paramagnetic state, the low-temperature scattering at
these positions must be entirely magnetic in origin. In addition,
there are no structure superlattice
reflections around $(H,K,L)=(\pm(2m+1)/2,\pm(2n+1)/2,0)$ where $m$, $n=0$, 1
as seen in the superconducting Nd$_{1.85}$Ce$_{0.15}$CuO$_4$.
The solid and dotted lines are Gaussian fits.
}
\end{figure}

Since the high temperature 
type-I phase has the same magnetic structure as type-III phase
but without the complication of the significantly polarized Nd moments, measurements 
there should provide information concerning the field
effect on only the Cu moments.
Figure 3 summarizes the magnetic field effect data taken in the type-I phase at 
100 K and in the paramagnetic state at 300 K.
Again, we find that a 7-T ${\bf B}||c$-axis field neither  
induces new magnetic order at $(1/2,1/2,0)$ and $(1/2,0,0)$ nor enhances 
the AF $(1/2,3/2,0)$ peak present in type-I phase 
(Figs. 3a-c). The enhancement of the $(1,1,0)$ Bragg
intensity is still present at 100 K, but is too small to observe in the paramagnetic state at 300 K. 
The absence of peaks around the $(1/2,3/2,0)$ and $(1/2,1/2,0)$ positions at 300 K 
indicates that the low-temperature
reflections at these positions are entirely magnetic in origin.

\begin{figure}
\includegraphics[keepaspectratio=true,width=0.8\columnwidth,clip]{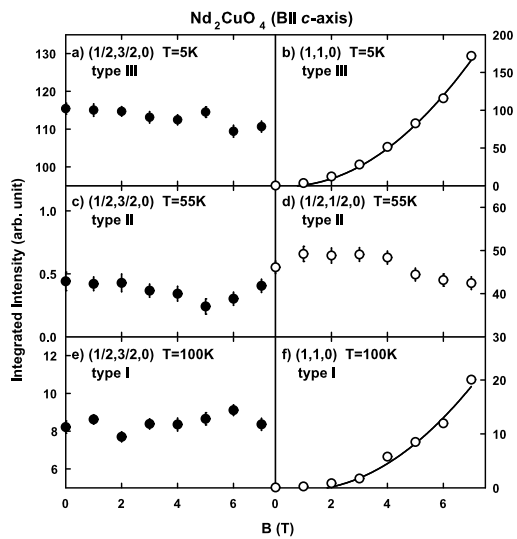}
\caption{Effect of a ${\bf B}||c$-axis field on the integrated intensity
of AF and FM Bragg reflections in all AF phases of Nd$_{2}$CuO$_{4}$. 
The field dependence of the integrated intensity of (a) $(1/2,3/2,0)$ and
(b)$(1,1,0)$ at 5 K in type-III phase; (c) $(1/2,3/2,0)$ and (d) 
$(1/2,1/2,0)$ at 55 K in type-II phase; (e) $(1/2,3/2,0)$ and (f) $(1,1,0)$
at 100 K in type-I phase. The quadratic field-dependent FM intensity is
clearly evident in (b) and (f), suggesting that field-induced FM moments
increase linearly with increasing field. 
}
\end{figure}

In Fig. 4, we summarize the effect of magnetic fields 
on AF and FM ordering on Nd$_{2}$CuO$_{4}$.
At 5 K in the type-III phase, 
the integrated intensity of the residual AF $(1/2,3/2,0)$ peak 
decreases slightly with increasing field (Fig. 4a), 
while the field-induced FM $(1,1,0)$ intensity increases 
quadratically with increasing field (Fig. 4b). 
The decreasing $(1/2,3/2,0)$ intensity with field suggests 
a small canting of the Cu(Nd) moments towards the field direction.
The quadratic increase in the $(1,1,0)$ intensity indicates that 
the field-induced Cu(Nd) FM moments increase linearly 
with increasing field, as the measured neutron intensity 
is proportional to square of the magnetic moment. 
At 55 K in the type-II phase, we find that 
while the AF $(1/2,3/2,0)$ and $(1/2,1/2,0)$ reflections 
change negligibly with field (Figs. 4c and 4d), 
the $(1,1,0)$ intensity again increases quadratically 
with increasing field (not shown).

Figs. 4e and 4f show the data obtained in the type-I phase at 100 K. 
As expected, the results are very similar to 
those of type-III phase except for the decreased
coefficient of the $(1,1,0)$ intensity quadratic curve 
compared to the type-III phase. 
Such a decrease is expected due to the reduced susceptibility 
of the Nd contribution to the
field-induced FM moments at higher temperatures.

Finally, we measure the
temperature dependence of the scattering at $(1/2,3/2,0)$, $(1,1,0)$, 
and $(1/2,1/2,0)$ under a ${\bf B}||c$-axis field 
to determine its influence across different AF phase transitions.
On application of a 7-T field, 
long-range FM ordering is induced below $\sim$250 K 
as seen by the added magnetic intensity 
to the $(1,1,0)$ structural Bragg peak intensity (Fig. 1d). 
A 7-T field thus induces FM moments on Cu(Nd) sites 
not far below $T_{N1}$. On the other hand, 
there is very little intensity change between 0 and 7-T 
at the $(1/2,3/2,0)$ and $(1/2,1/2,0)$ positions 
across $T_{N2}$ and $T_{N3}$ (Figs. 1c and 1e). 
Therefore, it becomes clear that 
antiferromagnetism in all three phases of Nd$_{2}$CuO$_{4}$
and transitions across them are not strongly affected
by the applied magnetic field.

\section{Results on as-grown nonsuperconducting 
N\lowercase{d}$_{1.85}$C\lowercase{e}$_{0.15}$C\lowercase{u}O$_{4}$}

Although our results show conclusively that a 7-T magnetic
field has no effect on the long-range AF order in all phases of 
Nd$_2$CuO$_4$, one still needs to determine the magnetic field effect
on as-grown nonsuperconducting Nd$_{1.85}$Ce$_{0.15}$CuO$_{4}$ because 
Ce-doping may influence the magnetic response of the system to a $c$-axis
aligned field. Consistent with earlier work on as-grown nonsuperconducting
Nd$_{1.85}$Ce$_{0.15}$CuO$_{4}$ \cite{yamada}, we find that the system
orders antiferromagnetically with a type-I/III structure. In addition, the 
as-grown samples are pure Nd$_{1.85}$Ce$_{0.15}$CuO$_{4}$ and 
have no known impurity phases.

\begin{figure}
\includegraphics[keepaspectratio=true,width=0.8\columnwidth,clip]{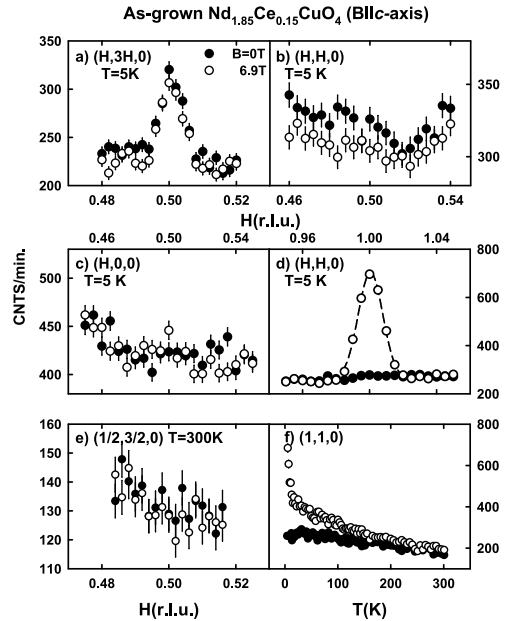}
\caption{Effect of a ${\bf B}||c$-axis field on the AF peaks (half integer)
and field-induced FM peaks in as-grown nonsuperconducting Nd$_{1.85}$Ce$_{0.15}$CuO$_{4}$. 
Scans around (a) AF Bragg reflection 
$(1/2,3/2,0)$, (b) $(1/2,1/2,0)$, 
(c) $(1/2,0,0)$, and (d) the FM Bragg peak $(1,1,0)$
at 5 K.
(e) Scattering in the paramagnetic state at 300 K around $(1/2,3/2,0)$.
(f) Temperature dependence of the scattering at
the FM Bragg peak $(1,1,0)$ position. The closed and open circles 
represent identical scans at zero and 7-T field,
respectively.}
\end{figure}

Figure 5 summarizes 
the effect of a 7-T $c$-axis aligned field to the AF structure of
as-grown Nd$_{1.85}$Ce$_{0.15}$CuO$_{4}$. At zero field and 5 K, we find the AF 
peak at $(1/2,3/2,0)$ (Fig. 5a),
no magnetic scattering at $(1/2,1/2,0)$ (Fig. 5b) and
$(1/2,0,0)$ (Fig. 5c), consistent with the type-I/III structure (see Figs. 2 and 3).
On application of a 7-T $c$-axis aligned field, the scattering
remains unchanged at the AF position $(1/2,3/2,0)$ (Fig. 5a) but is  
enhanced dramatically at the FM position $(1,1,0)$ (Fig. 5d). In addition, we find no
evidence of field-induced peaks at $(1/2,1/2,0)$ (Fig. 5b) 
and $(1/2,0,0)$ (Fig. 5c). On warming the system to room temperature,
the AF $(1/2,3/2,0)$ peak disappears, thus indicating that  
the low-temperature intensity is entirely magnetic in origin. Since the 
$(1/2,3/2,0)$ reflection has the same temperature dependence as $(1/2,1/2,3)$ \cite{jeff},
the absence of a field-induced effect at $(1/2,3/2,0)$ is direct 
evidence of no field-induced effect at $(1/2,1/2,3)$ in as-grown 
nonsuperconducting Nd$_{1.85}$Ce$_{0.15}$CuO$_{4}$. Therefore, we conclude 
that a 7-T $c$-axis aligned magnetic field has negligible effect on the
AF order of the system.  

\section{Results on superconducting 
N\lowercase{d}$_{1.85}$C\lowercase{e}$_{0.15}$C\lowercase{u}O$_{4}$ and 
the effect of cubic (N\lowercase{d},C\lowercase{e})$_2$O$_3$ impurity phase}

We begin this section by summarizing the effect of a $c$-axis aligned 
magnetic field on magnetic scattering of superconducting 
Nd$_{1.85}$Ce$_{0.15}$CuO$_{4}$. 
Below $T_c$, such a field induces magnetic 
scattering at $(\pm (2m+1)/2, \pm (2n+1)/2, 0)$, $(\pm (2m+1)/2, \pm n, 0)$, and 
$(\pm m, \pm (2n+1)/2, 0)$ with $m, n = 0$, 1, 2 \cite{kang}.  
Figure 6 shows our survey scans at various places in reciprocal space. At zero field, 
Nd$_{1.85}$Ce$_{0.15}$CuO$_{4}$ orders antiferromagnetically in the type-III structure
and has magnetic peaks at $(\pm 1/2,\pm 3/2,0)$ and $(\pm 3/2,\pm 1/2,0)$. 
Inspection of Figure 6 reveals that in addition to the magnetic $(3/2,1/2,0)$ peak (Fig. 6b), 
there are structural reflections at most superlattice positions in the $a$-$b$ plane.

\begin{figure}
\includegraphics[keepaspectratio=true,width=0.8\columnwidth,clip]{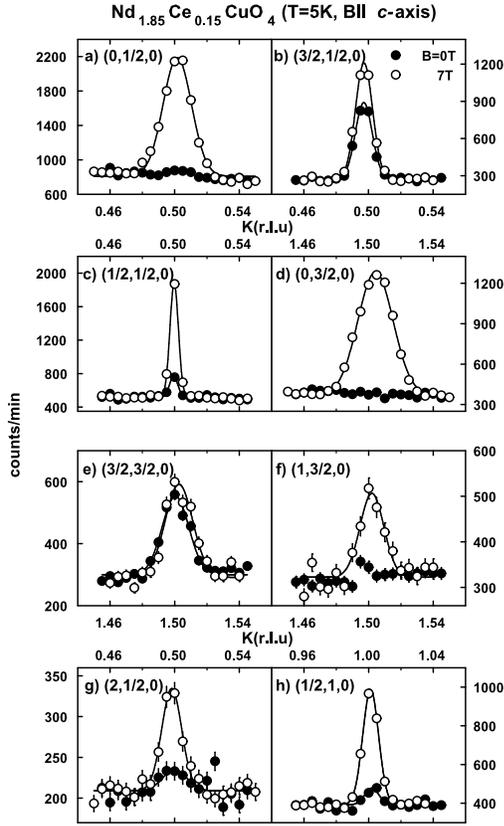}
\caption{Effect of a ${\bf B}||c$-axis field on the integrated intensity
of AF Bragg reflections and superlattice positions at $T=5$ K in the $(H,K,0)$ 
scattering plane of superconducting Nd$_{1.85}$Ce$_{0.15}$CuO$_{4}$. 
The field dependence of the integrated intensity of (a) $(0,1/2,0)$;
(b)$(3/2,1/2,0)$ ; (c) $(1/2,1/2,0)$; (d) $(0,3/2,0)$; 
(e) $(3/2,3/2,0)$; (f) $(1,3/2,0)$; (g) $(2,1/2,0)$; and 
(h) $(1/2,1,0)$. The filled circles represent 0-T data while the open 
circles are identical scans at 7-T. The scattering at $(3/2,3/2,0)$
is mostly from the epitaxial cubic (Nd,Ce)$_2$O$_3$ (see below) and has 
weak field-induced effect up to 7-T. Note the observation of 
clear superlattice peaks at $(1/2,0,0)$, $(1/2,1,0)$, $(1,3/2,0)$
positions disallowed by cubic (Nd,Ce)$_2$O$_3$. The solid lines are Gaussian fits.
}
\end{figure}

To demonstrate that the field-induced effect in ref. \cite{kang} and 
Fig. 6 indeed arises from the suppression of superconductivity, we not only need to show
that similar field-induced effects are not there in the parent compound and
as-grown nonsuperconducting Nd$_{1.85}$Ce$_{0.15}$CuO$_{4}$, but we also have to rule out 
other spurious effects. One possible spurious effect is the formation of 
cubic (Nd,Ce)$_2$O$_3$ as an impurity phase due to the partial decomposition of 
Nd$_{1.85}$Ce$_{0.15}$CuO$_{4}$ crystal during the annealing process \cite{meng,brauer}.
In general, impurity phases resulting from a heat treatment procedure should create powder lines
unrelated to the original underlying lattice. However, the cubic (Nd,Ce)$_2$O$_3$ stabilizes as 
an oriented crystalline lattice in the crystal because of its close lattice 
parameter matching to the tetragonal planes of Nd$_{1.85}$Ce$_{0.15}$CuO$_{4}$
($a=3.945$ \AA\ and $a_{NO}\approx 2\sqrt{2}a$). To distinguish the cubic (Nd,Ce)$_2$O$_3$ from
Nd$_{1.85}$Ce$_{0.15}$CuO$_{4}$, one needs to perform scans along the $c$-axis direction as
the lattice parameter of the former ($a_{NO}=11.072$ \AA) is significantly different from that of 
the latter ($c=12.07$ \AA). 
Table II summarizes the Miller indexes of the nonzero structural factors for the cubic 
(Nd,Ce)$_2$O$_3$ assuming the Mn$_2$O$_3$ structure type. For comparison, we also label their 
corresponding Miller indexes in the tetragonal unit cells of Nd$_{1.85}$Ce$_{0.15}$CuO$_4$.

\begin{table}
\caption{ The calculated lattice $d$-spacings, structural factors, and Miller indexes 
for the cubic (Nd,Ce)$_2$O$_3$ (NO)assuming $a_{NO}=11.072$ \AA. For comparison with experiments,
we also label their 
corresponding Miller indexes in the tetragonal unit cell of 
Nd$_{1.85}$Ce$_{0.15}$CuO$_4$ (NCCO) along
the $[1/2,1/2,L]$, $[1/2,0,L]$, and $[3/2,3/2,L]$ directions. }
\begin{ruledtabular}
\begin{tabular}{cccc}
NO$(H,K,L)$ &$d$-spacing(\AA)&$|F(H,K,L)|$ & NCCO$(H,K,L)$ \\
\hline
$(0,0,2)$ & 5.539     & 11.14  & $(0,0,2.178)$ \\
$(0,2,0)$ & 5.539     & 11.14  & $(0.504,0.504,0)$ \\
$(0,2,2)$ & 3.917     & 7.18  &  $(1/2,1/2,2.178)$ \\
$(0,2,4)$ & 2.477     & 16.85  & $(1/2,1/2,4.351)$ \\
$(1,1,2)$ & 4.523     & 40.11  & $(1/2,0,2.176)$ \\
$(1,1,4)$ & 2.611     & 22.68  & $(1/2,0,4.35)$ \\
$(0,6,0)$ & 1.846     & 27.92  & $(1.512,1.512,0)$ \\
$(0,6,2)$ & 1.751     & 48.51  & $(3/2,3/2,2.202)$ \\
$(0,6,4)$ & 1.536     & 56.21  & $(3/2,3/2,4.363)$ \\
\end{tabular}
\end{ruledtabular}
\end{table}

To estimate the fractional volume of the cubic (Nd,Ce)$_2$O$_3$ in our superconducting 
Nd$_{1.85}$Ce$_{0.15}$CuO$_4$, we aligned the crystal in the $(H,0,L)$ and 
$(H,H,L)$ zones and performed $c$-axis scans along the $[1/2,0,L]$ and 
$[3/2,3/2,L]$ directions, respectively, at room temperature. Figure 7 plots the
outcome of the experiment. Along the $[1/2,0,L]$ direction (Fig. 7a), sharp 
resolution-limited Bragg peaks corresponding to the cubic $(1,1,\pm 2)$ 
and $(1,1,\pm 4)$ reflections are observed at $(1/2,0,\pm2.176)$ and 
$(1/2,0,\pm4.35)$ in the tetragonal Miller indexes of 
Nd$_{1.85}$Ce$_{0.15}$CuO$_4$, respectively (Table II). 
Along the $[3/2,3/2,L]$ direction, the cubic (Nd,Ce)$_2$O$_3$ 
$(0,6,0)$, $(0,6,2)$, and $(0,6,4)$ peaks are observed at the expected places (Fig. 7b).
The observation of sharp Bragg peaks from (Nd,Ce)$_2$O$_3$ along the 
$c$-axis and in the CuO$_2$ plane indicates that cubic (Nd,Ce)$_2$O$_3$ forms 
three-dimensional long-range order in the matrix of Nd$_{1.85}$Ce$_{0.15}$CuO$_4$.   
By comparing the large (Nd,Ce)$_2$O$_3$ $(2,2,2)$ Bragg peak ($\sim$8700 counts/minute) 
with the very weak $(1,0,1)$ ($\sim$21160 counts/minute) reflection of 
Nd$_{1.85}$Ce$_{0.15}$CuO$_4$, we estimate that (Nd,Ce)$_2$O$_3$ has a volume fraction 
1.0$\times 10^{-5}$. Alternatively, if we use the very strong  
$(2,0,0)$ ($\sim$1.08$\times 10^8$ counts/minute) reflection of Nd$_{1.85}$Ce$_{0.15}$CuO$_4$,
we find a volume fraction of 2.0$\times 10^{-3}$ for (Nd,Ce)$_2$O$_3$. However, the very strong
fundamental peaks are severely extinction limited, and this overestimates the 
(Nd,Ce)$_2$O$_3$ volume fraction (just as the ordered Cu moment is overestimated in Table I 
using the $(2,0,0)$ reflection). Therefore, the estimate using the weak  
Nd$_{1.85}$Ce$_{0.15}$CuO$_4$ structural peaks is more reliable.

For the cubic (Nd,Ce)$_2$O$_3$ with the $Ia3$ space group symmetry, 
structure factor calculations 
show vanishing intensity at $(1/2,0,0)$ and equivalent positions. 
Although the absence of a sharp Bragg peak 
along the $[1/2,0,L]$ direction at $L\approx0$ 
confirms the structure factor calculation, 
the $(1/2,0,0)$ peak that 
is sharp along the $[H,0,0]$ direction (see Fig. 2c, Ref. \cite{kang} and Fig. 6a, section V)
but diffusive along the $[1/2,0,L]$ direction (Fig. 7a)
is reminiscent of the superlattice reflection seen at $(1/2,1/2,0)$ \cite{kurahashi}. 
Since these diffuse superlattice reflections at (1/2,0,0) and (1/2,1/2,0)
have no magnetic field dependence at 5 K 
and are not related to the cubic (Nd,Ce)$_2$O$_3$, 
they must be associated with the formation of a quasi two-dimensional 
lattice distortion necessary for Nd$_{1.85}$Ce$_{0.15}$CuO$_4$ to become
superconducting. Work is currently underway to determine the microscopic 
origin of the lattice distortion.

\begin{figure}
\includegraphics[keepaspectratio=true,width=0.8\columnwidth,clip]{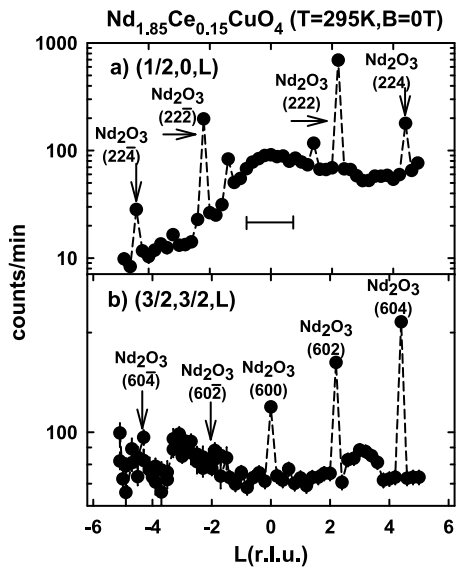}
\caption{Room temperature scans to determine the cubic (Nd,Ce)$_2$O$_3$ 
impurity phase and $c$-axis modulation of the structural superlattice reflections 
in superconducting Nd$_{1.85}$Ce$_{0.15}$CuO$_4$.  
(a) $L$-scan along the $[1/2,0,L]$ direction. In addition to the well-marked 
cubic (Nd,Ce)$_2$O$_3$ peaks, a broad diffusive peak with 
a full-width at half maximum of $\Delta L=1.1$ rlu is observed at $L=0$. Such 
diffusive peak is similar to the superlattice reflections reported in
Ref. \cite{kurahashi} at $(1/2,1/2,0)$. We also note that 
part of the peak intensity at $L=0$ arises from the small detector angles 
and the neutron absorption of the long Nd$_{1.85}$Ce$_{0.15}$CuO$_4$ crystal. 
(b) $L$-scan along the $[3/2,3/2,L]$ 
direction with (Nd,Ce)$_2$O$_3$ peaks marked by arrows. Since no broad diffusive
peak is found at $L=0$, most of the intensity at $(3/2,3/2,0)$ in Fig. 6e 
is due to (Nd,Ce)$_2$O$_3$. 
}
\end{figure}

The identification of the epitaxial cubic (Nd,Ce)$_2$O$_3$ with lattice parameter 
close to that of the superconducting Nd$_{1.85}$Ce$_{0.15}$CuO$_4$  
raises the important question concerning the possible role of this impurity phase
in the observed field-induced effect \cite{kang}, as the rare-earth magnetic ion Nd$^{3+}$ 
in (Nd,Ce)$_2$O$_3$ will be polarized by the applied field. In general, the  
rare-earth oxides such as Nd$_2$O$_3$ and Er$_2$O$_3$ have the 
bixbyite structure with 32 rare-earth ions in a cubic
unit cell and order antiferromagnetically
at low temperature \cite{moon}. Since scattering at all half integer positions 
[except $(1/2,3/2,0)$ from Nd$_{1.85}$Ce$_{0.15}$CuO$_4$] is temperature independent
above 5 K \cite{kang}, it is safe to assume that the (Nd,Ce)$_2$O$_3$ impurity is
in the paramagnetic state at this temperature.

In the paramagnetic state of (Nd,Ce)$_2$O$_3$,
a field will induce a net moment given by a Brillouin function, 
and the field-induced moment should saturate in the high field limit.
This is in clear contrast to our observation where the scattering first 
increases with field, and then decreases at higher fields at 5 K \cite{kang}.  
Of course, at sufficiently low temperatures where (Nd,Ce)$_2$O$_3$ and/or  
Nd in Nd$_{1.85}$Ce$_{0.15}$CuO$_4$ ($T_{N}\approx 1.2$ K) 
spontaneously order \cite{lynnprb}, an applied field will rotate the ordered 
AF moment along the field direction and therefore suppress the AF intensity. 
The results we report in Ref. \cite{kang} carefully avoided the regime 
of spontaneous magnetic order for both (Nd,Ce)$_2$O$_3$ 
and superconducting Nd$_{1.85}$Ce$_{0.15}$CuO$_4$ \cite{lynnprb}
by measuring spectra above 5 K.

Since the cubic (Nd,Ce)$_2$O$_3$ impurity phase has almost the same lattice parameter 
as Nd$_{1.85}$Ce$_{0.15}$CuO$_4$ in the $a$-$b$ plane, measurements at 
$L=0$ could be ambiguous as the scattering could originate from either (Nd,Ce)$_2$O$_3$ or 
Nd$_{1.85}$Ce$_{0.15}$CuO$_4$. The experimental resolution of this ambiguity is 
straightforward, 
measurements simply need to be made at finite $L$,
where the Nd$_{1.85}$Ce$_{0.15}$CuO$_4$ peaks are not coincident with (Nd,Ce)$_2$O$_3$.  
To accomplish this, we aligned the crystal
in the $(H,H,L)$ zone inside the HM2
4-T horizontal field magnet at HMI. In this geometry, 
we can probe the $L$-dependence of the scattering 
while keeping the field along the $c$-axis. 
Figure 8 summarizes the outcome of the experiment. 
At zero-field and 5 K, the $[1/2,1/2,L]$ scan
shows well-defined peaks associated with the residual 
AF order of Nd$_{1.85}$Ce$_{0.15}$CuO$_4$
at $(1/2,1/2,3)$ and $(1/2,1/2,5)$. In addition, we find the $(0,2,4)$ reflection of 
the cubic (Nd,Ce)$_2$O$_3$ and the $(1,1,1)$ powder 
peak of the aluminum sample holder (Fig. 8a and Table II). 
When $c$-axis aligned fields are applied, the residual AF $(1/2,1/2,3)$ peak enhances  
systematically with increasing field (Fig. 8b)
while the (Nd,Ce)$_2$O$_3$ $(0,2,4)$ (Figs. 8a and c) and the aluminum $(1,1,1)$ 
reflections (Fig. 8a) are not affected.

Figures 8d and e show the $[H,H,3]$ scan and temperature dependence 
of the scattering at the $(1/2,1/2,3)$, respectively. Clear field-induced enhancements are observed
below $T_c$, consistent with the data in the $(H,K,0)$ plane \cite{kang}.
We note that a 2-T field parallel to the
CuO$_2$ along the $[1,-1,0]$ direction induces a spin-flop transition and 
suppresses the intensity at $(1/2,1/2,3)$ (See Fig. 4 in ref. \cite{kang}). 
Therefore, the $(1/2,1/2,3)$ peak shows an induced AF component 
when the field is along the $c$-axis and superconductivity is strongly suppressed, 
but not when it is in the $a$-$b$ plane and superconductivity is only weakly 
affected \cite{kang}.  We also note that the qualitatively different behavior observed 
for ${\bf B}||ab$-plane versus ${\bf B}||c$-axis for $(1/2,1/2,3)$
directly violates the cubic symmetry of (Nd,Ce)$_2$O$_3$.

\begin{figure}
\includegraphics[keepaspectratio=true,width=0.8\columnwidth,clip]{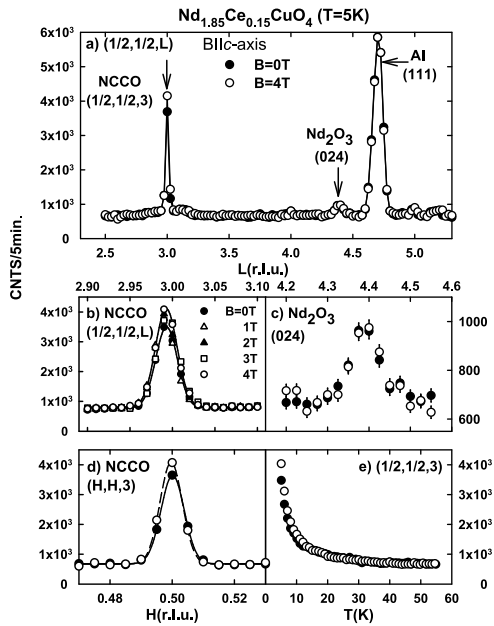}
\caption{Effect of a ${\bf B}||c$-axis field on the integrated intensity
of AF Bragg reflections and the epitaxy Nd$_2$O$_3$ at $T=5$ K in the $(H,H,L)$ 
scattering plane using the HM-2 horizontal field magnet at HMI. 
The filled circles represent 0-T data while the open 
circles are identical scans at 4-T.
(a) The $[1/2,1/2,L]$ scan at 0-T and 4-T with superconducting Nd$_{1.85}$Ce$_{0.15}$CuO$_4$,
(Nd,Ce)$_2$O$_3$, and aluminum peaks marked by the arrows.
(b) Detailed scans along the $[1/2,1/2,L]$ direction around $(1/2,1/2,3)$ reflection
at various fields. (c) Detailed scans around the cubic (Nd,Ce)$_2$O$_3$ $(0,2,4)$ peak 
at 0-T and 4-T. There is no observable field-induced effect at 4 T. 
(d) The $[H,H,3]$ scan around $(1/2,1/2,3)$ reflection
at 0 T and 4 T. (e) The temperature dependence of the scattering at $(1/2,1/2,3)$ at
0-T and 4-T. The solid and dotted lines in (b) and (d) are Gaussian fits.
}
\end{figure}

Figure 9 compares 
the temperature and field dependence of the field-induced scattering
at $(1/2,3/2,0)$ (Figs. 9a and c) and $(1/2,1/2,3)$ (Figs. 9b and d).
The remarkable similarity of the field response in these reflections
suggests that they must originate from the same physical process. 
Considering that a 7-T $c$-axis aligned field 
has no effect on $(1/2,3/2,0)$
in as-grown nonsuperconducting Nd$_{1.85}$Ce$_{0.15}$CuO$_4$ (see section IV), 
we conclude that field-induced AF order at $(1/2,3/2,0)$ and $(1/2,1/2,3)$ 
in the Figure can only result from the suppression of superconductivity.
Furthermore, the new data indicate that the field-induced 
enhancement forms three-dimensional
long-range AF order in superconducting Nd$_{1.85}$Ce$_{0.15}$CuO$_4$.

\begin{figure}
\includegraphics[keepaspectratio=true,width=0.8\columnwidth,clip]{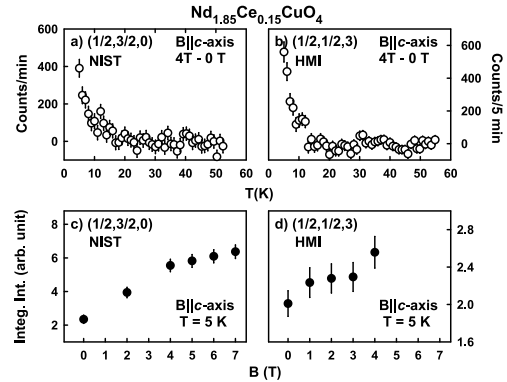}
\caption{Comparison of the field-induced effect for superconducting 
Nd$_{1.85}$Ce$_{0.15}$CuO$_4$ at $(1/2,3/2,0)$ and $(1/2,1/2,3)$.
While the data at $(1/2,3/2,0)$ are from Ref. \cite{kang}, the results 
at $(1/2,1/2,3)$ are new. The temperature dependence
of the difference between 4 T and 0 T at (a) $(1/2,3/2,0)$ and (b) $(1/2,1/2,3)$.
The field dependence of the integrated intensity at (c) $(1/2,3/2,0)$ and (d) $(1/2,1/2,3)$.}
\end{figure}

Although our horizontal field measurements conclusively demonstrate that a $c$-axis aligned 
magnetic field
enhances the residual AF order in superconducting Nd$_{1.85}$Ce$_{0.15}$CuO$_4$, it is also 
important to determine the field-induced effect of the (Nd,Ce)$_2$O$_3$ impurity phase.
Because (Nd,Ce)$_2$O$_3$ forms a three-dimensional long-range 
ordered cubic lattice (Fig. 7), 
its field-induced effect should be isotropic for fields along (Nd,Ce)$_2$O$_3$ 
$[2,0,0]$ and $[0,0,2]$ directions. In the notation of Nd$_{1.85}$Ce$_{0.15}$CuO$_4$ Miller 
indexes (Table II), 
these are along $[1,-1,0]$ (${\bf B}||ab$-plane) and $[0,0,1]$ (${\bf B}||c$-axis) 
directions, respectively. As superconductivity is strongly suppressed for a ${\bf B}||c$-axis
field but much less affected by the same field in the $ab$-plane, measurements of
field directional anisotropy will establish the influence of (Nd,Ce)$_2$O$_3$ to 
the observed field effect in Nd$_{1.85}$Ce$_{0.15}$CuO$_4$ \cite{kang}.

Figure 10 summaries the outcome of such experiment on BT-2 using the $same$ crystal
of superconducting Nd$_{1.85}$Ce$_{0.15}$CuO$_4$ in two different field geometries. We first
describe measurements in the $(H,H,L)$ zone, where the applied vertical field 
is along the $[1,-1,0]$ direction of Nd$_{1.85}$Ce$_{0.15}$CuO$_4$ and $[2,0,0]$ direction
of (Nd,Ce)$_2$O$_3$.
In this geometry, we can probe the field-induced effect on 
$(1/2,1/2,0)$ and $(0,0,2.2)$ without much affecting the superconductivity. 
While scattering at $(1/2,1/2,0)$ may originate from
either Nd$_{1.85}$Ce$_{0.15}$CuO$_4$ or (Nd,Ce)$_2$O$_3$, 
$(0,0,2.2)$ is exclusively associated with the $(0,0,2)$ reflection
of (Nd,Ce)$_2$O$_3$ (Table II). 
Figures 10a and b show the outcome of the experiment along the 
$[H,H,0]$ and $[0,0,L]$ directions at 5 K. 
Clear field-induced effects are seen at $(0,0,2.2)$ (Fig. 10b), indicating that 
(Nd,Ce)$_2$O$_3$ can indeed be polarized by the applied field. Similar measurements 
on $(0,0,2.2)$ with field aligned along the $[1,-1,0]$ direction
of (Nd,Ce)$_2$O$_3$ show a weaker field-induced
effect, thus suggesting that its easy axis is along the $[2,0,0]$ direction.  
If we ignore the contribution of the superlattice structure to $(1/2,1/2,0)$
and assume that the scattering there is due entirely to 
(Nd,Ce)$_2$O$_3$ (Table II), its integrated intensity should be
identical to that at $(0,0,2.2)$ because of the cubic symmetry of (Nd,Ce)$_2$O$_3$.
By normalizing the zero field intensity at $(0,0,2.2)$ to that at $(1/2,1/2,0)$,
we can compare the field-induced effect at these two equivalent positions
for (Nd,Ce)$_2$O$_3$. 
Since the field-induced effect at $(1/2,1/2,0)$ (Fig. 10a) 
is only about 25\% larger that at $(0,0,2.2)$ (Fig. 10b), we conclude that 
the field-induced intensity at $(1/2,1/2,0)$ is mostly due to the polarization
of (Nd,Ce)$_2$O$_3$ in this field geometry.

\begin{figure}
\includegraphics[keepaspectratio=true,width=0.8\columnwidth,clip]{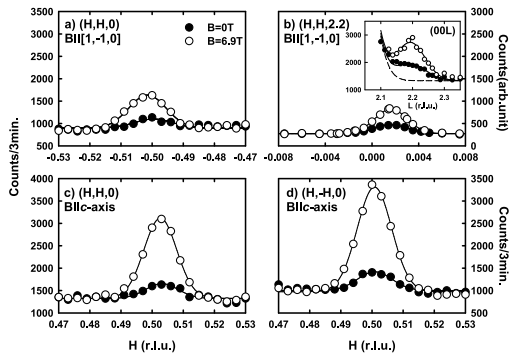}
\caption{Comparison of the field-induced effect at 5 K for superconducting 
Nd$_{1.85}$Ce$_{0.15}$CuO$_4$ at $(1/2,1/2,0)$ and $(0,0,2.2)$ in the ${\bf B}||ab$-plane geometry 
with that at $(1/2,1/2,0)$ and $(1/2,-1/2,0)$ in the ${\bf B}||c$-axis geometry.
Scans through (a) $(1/2,1/2,0)$ and (b) $(0,0,2.2)$ at 5 K for field along the $[1,-1,0]$ axis.
Since $(0,0,2.2)$ from (Nd,Ce)$_2$O$_3$ is very close to the strong
$(0,0,2)$ nuclear Bragg peak from Nd$_{1.85}$Ce$_{0.15}$CuO$_4$, the radial scan
in the inset of (b) shows sloped background. We determine the background scattering 
at $(0,0,2.2)$ by fitting a Gaussian with fixed width to $(0,0,2)$ (dashed line
in the inset). This is confirmed by the transverse scan across $(0,0,2.2)$.    
Similar scans through (c) $(1/2,1/2,0)$ and (d) $(1/2,-1/2,0)$ for 
field along the $[0,0,1]$ direction. 
The ratios of integrated intensities
of $(0,0,2.2)$ and $(1/2,1/2,0)$ between 
6.9 T and 0 T ($I$(6.9 T)$/I$(0 T))
are 2.7 and 3.3, respectively, for the ${\bf B}||ab$-plane field.
In comparison, the ratio  
is 5.7 for both $(1/2,1/2,0)$ and $(1/2,-1/2,0)$ for the $c$-axis aligned field. 
In ref. \cite{kang} and Fig. 6,
the ratios of $I$(7 T)$/I$(0 T) at $(1/2,1/2,0)$ are 6.8 and 5.5, respectively,
in the ${\bf B}||c$-axis geometry.}
\end{figure}

In Figures 10c and d, we plot the field-induced effect for the 
${\bf B}||c$-axis experiment
at two equivalent reflections $(1/2,1/2,0)$
 and $(1/2,-1/2,0)$. While the magnitude of
field-induced effect is consistent with Ref. \cite{kang}, 
they are twice as large as that of Figs. 10a and b. 
Since the cubic symmetry of (Nd,Ce)$_2$O$_3$ requires 
the same induced effect for fields along $[2,0,0]$ 
and $[0,0,2]$, the observation of a much larger field-induced 
effect in ${\bf B}||c$-axis geometry means the excess field-induced 
intensity must originate from the suppression of superconductivity.

Finally, we remark that one concern raised \cite{greven} was that the finite 
$L$ behavior we observe  
might not be intrinsic to superconducting Nd$_{1.85}$Ce$_{0.15}$CuO$_4$, but rather 
it is somehow induced by the magnetic coupling to the (Nd,Ce)$_2$O$_3$ impurity phase.  
However, (Nd,Ce)$_2$O$_3$ has a very weak exchange interaction and thus 
orders at very low temperature. It is difficult to see how (Nd,Ce)$_2$O$_3$ could dominate the 
Nd$_{1.85}$Ce$_{0.15}$CuO$_4$
physics, particularly when it only constitutes a 
small ($\sim$10$^{-5}$) volume fraction of the crystals
in superconducting Nd$_{1.85}$Ce$_{0.15}$CuO$_4$. 
On the other hand, the Nd magnetic structure
in Nd$_{1.85}$Ce$_{0.15}$CuO$_4$ has 
the same symmetry as the Cu spin configuration and thus is maximally coupled to the Cu spins \cite{lynnprb}. 
Even in this case, the perturbation of the Nd order parameter by the Cu spins is small. Therefore, 
while the (Nd,Ce)$_2$O$_3$ ordering may be
induced by being in contact with bulk
Nd$_{1.85}$Ce$_{0.15}$CuO$_4$, it is highly 
improbable that the field-induced magnetic scattering 
at $(1/2,1/2,3)$ could be induced by the (Nd,Ce)$_2$O$_3$ impurity.

\section{Summary and Conclusions}

We have investigated the effect of a ${\bf B}||c$-axis field in all phases
of Nd$_{2}$CuO$_{4}$ and in as-grown nonsuperconducting and 
superconducting Nd$_{1.85}$Ce$_{0.15}$CuO$_4$. 
At zero-field, Cu spins in Nd$_{2}$CuO$_{4}$ form noncollinear structures 
because of the coupling between Cu$^{2+}$ and Nd$^{3+}$. 
Such a magnetic interaction also creates a small in-plane spin-wave gap 
$\Delta_0$ at ${\bf B}=0$.
For a magnetic field aligned parallel to the CuO$_2$ plane, 
Cu spins transform from a noncollinear to collinear structure in a spin-flop phase 
transition with a critical field less than 2-T \cite{jeff,skanthakumar1,skanthakumar2}.
Such a spin-flop transition occurs because when the 
magnetic field associated with the Zeeman energy ($g\mu B_c$)
equals to $\Delta_0$, the net magnetic exchange interaction
vanishes and with it the noncollinear spin structure \cite{petitgrand}.

For a 7-T ${\bf B}||c$-axis field, our data clearly indicate
that the noncollinear AF spin structures in Nd$_{2}$CuO$_{4}$ are essentially 
unaffected for temperatures above 5 K. As a consequence, the zero-field in-plane spin-wave gap $\Delta_0$
and the magnetic exchange interaction must also remain unchanged in the field.
The large increase in the Cu(Nd) FM moments
suggests that the applied $c$-axis field only induces a canting of the
AF order. These results contrast significantly with that of superconducting
Nd$_{1.85}$Ce$_{0.15}$CuO$_4$, where the applied field induces a static,
long-range ordered AF state \cite{kang}. We demonstrate that the annealing process necessary
for superconductivity in Nd$_{1.85}$Ce$_{0.15}$CuO$_4$ 
also induces  
structural superlattice reflections at $(1/2,1/2,0)$ and $(1/2,0,0)$ positions. 
In addition, we confirm the presence of the cubic 
(Nd,Ce)$_2$O$_3$ as an impurity phase in superconducting Nd$_{1.85}$Ce$_{0.15}$CuO$_4$ 
following the annealing process \cite{meng}. Although the lattice parameter 
of the cubic (Nd,Ce)$_2$O$_3$ is very close to the in-plane lattice parameter of
Nd$_{1.85}$Ce$_{0.15}$CuO$_4$, most of the structural superlattice reflections 
in the $(H,K,0)$ plane are quasi two-dimensional and cannot be associated
with the three-dimensional cubic (Nd,Ce)$_2$O$_3$. By probing the $L$-dependence 
of the scattering with a $c$-axis aligned field, we show that 
the residual AF order in superconducting Nd$_{1.85}$Ce$_{0.15}$CuO$_4$ enhances with increasing field.
Such behavior is different from the field effect on as-grown nonsuperconducting 
Nd$_{1.85}$Ce$_{0.15}$CuO$_4$, where the long-range 
noncollinear AF order is essentially unaffected by a 7-T $c$-axis aligned field.  
By studying the anisotropy of field-induced effect, we determine the effect of 
the magnetic field on the cubic (Nd,Ce)$_2$O$_3$ and 
confirm that the results with the $c$-axis field in the $[H,K,0]$ plane are inconsistent
with the impurity phase. 
Combining these results with horizontal field experiments, 
we conclude that AF order is
induced in Nd$_{1.85}$Ce$_{0.15}$CuO$_4$ upon suppression of
superconductivity by a $c$-axis aligned magnetic field.

\section{Acknowledgments}
We are grateful to Y. Ando, Henry Fu, S. A. Kivelson, D.-H. Lee, D. Mandrus, H. A. Mook, and S.-C. Zhang for 
helpful conversations. We also thank S. Larochelle and P. K. Mang for initially alerting us to the 
existence of the secondary Nd$_2$O$_3$ phase.
This work was supported by U.S. NSF DMR-0139882 and 
DOE under Contract No. DE-AC05-00OR22725.

\end{document}